\documentclass[letterpaper, 10 pt, conference]{ieeeconf}
\usepackage{cite}
\usepackage{amsmath,amssymb,amsfonts}
\usepackage{graphicx}

\usepackage{amsthm}
\usepackage{textcomp}
\theoremstyle{remark}
\newtheorem{remark}{Remark}
\theoremstyle{plain}

\newtheorem{assumption}{Assumption}
\newtheorem{definition}{Definition}
\usepackage{algorithm}	%For algorithms
\usepackage[noEnd]{algpseudocodex}
\usepackage{subfig}	%For subfigures
\usepackage{hyperref}	% for making links clickable
\usepackage{bm}
\usepackage{eurosym}

\allowdisplaybreaks

\usepackage{tikz}
\usetikzlibrary{spy}
\usepackage[utf8]{inputenc}
\usepackage{pgfplots}
\usepgfplotslibrary{groupplots,dateplot}
\usetikzlibrary{patterns,shapes.arrows}
\pgfplotsset{compat=newest}
\def\axisdefaultheight{110pt}

\def\BibTeX{{\rm B\kern-.05em{\sc i\kern-.025em b}\kern-.08em
    T\kern-.1667em\lower.7ex\hbox{E}\kern-.125emX}}
\begin{document}
\title{Integrated Online Monitoring and Adaptation of Process Model Predictive Controllers}
\author{Samuel Mallick, Laura Boca de Giuli, Alessio La Bella, Azita Dabiri, Bart De Schutter, and Riccardo Scattolini
\thanks{Samuel Mallick, Azita Dabiri and Bart De Schutter are with the Delft Center for Systems and Control, Delft University of Technology, The Netherlands (emails: \textsl{s.h.mallick@tudelft.nl}, \textsl{a.dabiri@tudelft.nl}, and \textsl{b.deschutter@tudelft.nl}.}
\thanks{Laura Boca de Giuli, Alessio La Bella, and Riccardo Scattolini are with the Department of Electronics, Information, and Bioengineering, Politecnico di Milano, Italy (e-mails: \textsl{laura.bocadegiuli@polimi.it}, \textsl{alessio.labella@polimi.it},  and
	\textsl{riccardo.scattolini@polimi.it}). }
\thanks{
	The work of Samuel Mallick, Azita Dabiri, and Bart De Schutter was funded by the European Research Council (ERC) under the European Union's Horizon 2020 research and innovation programme (Grant agreement No. 101018826 - ERC Advanced Grant CLariNet).
	The work of Laura Boca de Giuli, Alessio La Bella, and Riccardo Scattolini was carried out within the MICS (Made in Italy - Circular and Sustainable) Extended Partnership and received funding from Next-Generation EU (Italian PNRR - M4 C2, Invest 1.3 - D.D. 1551.11-10-2022, PE00000004). CUP MICS D43C22003120001.}
}

\maketitle

\begin{abstract}
	This paper addresses the design of an event-triggered, data-based, and performance-oriented adaptation method for model predictive control (MPC). The performance of such a strategy depends on the accuracy of the prediction model, which may require online adaptation to prevent performance degradation under changing conditions. Unlike existing methods that continuously update model and control parameters from data, potentially leading to catastrophic forgetting and unnecessary control modifications, we propose a novel approach based on statistical monitoring of closed-loop performance indicators. This framework enables the detection of performance degradation, and, when required, controller adaptation is performed via reinforcement learning and identification techniques. The proposed strategy is validated on a high-fidelity simulation of a district heating system benchmark.
\end{abstract}

% \begin{IEEEkeywords}
% 	Model predictive control, online learning, performance monitoring, reinforcement learning.
% \end{IEEEkeywords}

\section{Introduction}\label{sec:introduction}
Model predictive control (MPC) has become a state-of-the-art control paradigm for process control systems, largely due to its model-based predictive nature and constraint handling \cite{rawlings_model_2017}. However, even well-designed MPC controllers may not maintain high performance over time in the case of unexpected system variations, which may require model and/or control parameter retuning. 
In fact, a crucial component is the prediction model, the accuracy of which underpins the performance of the MPC controller.
Construction of accurate prediction models for complex processes, however, can be problematic, necessitating the derivation from physics principles of highly nonlinear relations, and often resulting in models that are computationally demanding.
To alleviate this issue, the use of data-driven prediction models has become a popular approach, where gray- or black-box functions are identified from data to approximate the process dynamics. % \cite{berberich2025overview}.
\textcolor{black}{However, as data-driven models are, in general, accurate only locally around specific operating regions, such models may not remain accurate as conditions change, leading to performance degradation.}
%Furthermore, for data-driven models, responding to underlying system changes is not trivial as, unlike physics-based models where a simple update of physical parameters suffices, a full re-identification may be required.

For these cases, approaches have been proposed for the online adaptation of MPC controllers.
In this context, a traditional methodology is to leverage system-identification (sysID) model adaptation, whereby the prediction model is re-identified using new data generated online \cite{zhu1998multivariable}.
However, identifying a new prediction model requires generating a substantial amount of sufficiently exciting data, which may be unreasonable online, potentially even requiring a pause of system operation.
Alternatively, there is a growing literature around performance-based controller adaptation of MPC controllers.
These methods take the viewpoint that the prediction model should be adapted to improve the closed-loop performance of the controller, rather than the prediction accuracy of the model \cite{piga2019performance}.
%with the two objectives not necessarily aligning when approximate data-driven models are used .
In this context, learning algorithms, such as reinforcement learning (RL) \cite{gros_data-driven_2020} or Bayesian optimization \cite{sorourifar2021data}, are used to adapt MPC components, including the prediction model, the cost, and the constraints, in order to improve closed-loop performance.
With the cost and the constraints as additional degrees of freedom, these methods offer greater flexibility in restoring controller performance compared to sysID methods.
However, in general, these methods are only effective when the required adaptation of the controller is minor, failing when the initial prediction model is completely inaccurate.
Furthermore, these methods are, in general, continuously adapting, %which, for critical process systems, is unacceptable from the perspective of system operators \cite{harris1999review}.
lacking a notion of when acceptable performance is lost or regained, i.e., control performance monitoring.

Beginning with the seminal work \cite{harris1989assessment}, many methods have been developed to assess the performance of a controller, relying principally on the computation of the so-called Harris index.
%, defined as the ratio between the `ideal' variance of the output achievable with a minimum variance regulator, and the real one, computable from data \cite{aastrom2012introduction}. 
%Notably, this approach uses closed-loop data without a-priori knowledge of process or disturbance model, except for the process time delay. 
The Harris index is, however, restricted to unconstrained, single-input, single-output, time-invariant systems and, in order to overcome these limitations, many algorithms have been developed \cite{zagrobelny2013quis, tyler1996performance}.
For MPC controllers, performance monitoring is a challenge, due to the fact that typically multi-input, multi-output nonlinear systems with constraints are considered \cite{zagrobelny2013quis, schafer2004multivariable, lee2010industrial}. 
In general, the existing approaches monitor performance considering either prediction model accuracy \cite{zhao_multi-step_2010}, incurred stage costs \cite{zagrobelny2013quis, schafer2004multivariable}, or system variable trajectories \cite{alghazzawi_model_2009}, each of which is only a limited description of controller performance.

In light of the above challenges, in this paper we propose a novel scheme that integrates statistical performance monitoring with online controller adaptation.
To this end, we propose a monitoring scheme to determine the acceptability of an MPC controller's performance online, measuring the statistical distance between actual and acceptable performance based on a suite of performance features.
We further propose to leverage performance-based controller adaptation as an initial, fast response to detected performance degradation, deploying sysID only when the former is unable to restore performance.  
%We note that the use of performance-based learning using a triggering mechanism is in contrast to its typical application \cite{gros_data-driven_2020, sorourifar2021data}, where the adaptation typically happens continuously until convergence.
% Here, in contrast, the adaptation is triggered on and off with the goal of maintaining acceptable performance.
%We then leverage this performance monitoring in an online adaptation scheme, where, when a loss of acceptable performance is detected, performance-based controller adaptation acts as a primary response, and where, when it is detected that acceptable performance is not restored, sysID acts as a fallback mechanism.
The result is a methodology that quantifies and measures the `acceptability' of an MPC controller, and adapts to loss of acceptable performance with triggered stages of online adaptation.
We validate the proposed approach on high-fidelity simulations of a district heating system, demonstrating how performance degradation can be identified and adequately responded to, under changing conditions. 

\textbf{Notation:} We use subscripts to denote the time step associated with a discrete time signal, e.g., $x_k$ for the signal $x$ at time step $k$, and denote the vector stacking a signal from time steps $k$ to $k^\prime > k$ as $\bm{x}_{k:k^\prime} = [x^\top_k, x^\top_{k+1}, \dots, x^\top_{k^\prime}]^\top$.
Furthermore, within an MPC context, $x_{\tau|k}$ denotes the predicted value of $x$ at time step $\tau$, with the prediction made at time step $k \leq \tau$.
We use a tilde to stack these predictions as $\tilde{\bm{x}}_{k:k^\prime} = [x^\top_{k|k}, x^\top_{k+1|k}, \dots, x^\top_{k^\prime|k}]^\top$.
Finally, $|\mathcal{S}|$ denotes the cardinality of a discrete set $\mathcal{S}$.

\section{Problem Setting}
Consider a process described by the discrete-time system 
\begin{equation}\label{eq:problem_setting-dynamics}
	x^+ = f(x, u, d, \phi), \: y = c(x, u, d, \phi),
\end{equation}
where the functions $f: \mathbb{R}^{n_x} \times \mathbb{R}^{n_u} \times \mathbb{R}^{n_d} \times \mathbb{R}^{n_\phi} \to \mathbb{R}^{n_x}$ and $c: \mathbb{R}^{n_x} \times \mathbb{R}^{n_u} \times \mathbb{R}^{n_d} \times \mathbb{R}^{n_\phi} \to \mathbb{R}^{n_y}$ respectively describe how the state $x \in \mathbb{R}^{n_x}$ and the output $y \in \mathbb{R}^{n_y}$ evolve for a given control input $u \in \mathbb{R}^{n_u}$ and disturbance $d \in \mathbb{R}^{n_d}$.
The parameter $\phi \in \mathbb{R}^{n_\phi}$ contains system parameters that govern the dynamics, e.g., matrices $A$ and $B$ for linear systems.
%or the thermal efficiencies and resistances in an energy system.

The system is controlled by an MPC controller with prediction horizon $N$, defined by the following nonlinear program:
\begin{equation}\label{eq:problem_setting-mpc}
	\begin{aligned}
		J(x_k, &\bm{d}_{k:k+N}, \theta_k) = \\
		&\min_{\substack{\tilde{\bm{x}}_{k:k+N+1},\\ \tilde{\bm{y}}_{k:k+N}, \\ \tilde{\bm{u}}_{k:k+N}}} \: L(\tilde{\bm{x}}_{k:k+N+1}, \tilde{\bm{y}}_{k:k+N}, \tilde{\bm{u}}_{k:k+N}, \bm{d}_{k:k+N}, \theta_k) \\
		\text{s.t.} \quad &\tilde{\bm{x}}_{k+1:k+N+1} = \tilde{f}(x_k, \tilde{\bm{u}}_{k:k+N}, \bm{d}_{k:k+N}, \theta_k)  \\
		&\tilde{\bm{y}}_{k:k+N} = \tilde{c}(\tilde{\bm{x}}_{k:k+N}, \tilde{\bm{u}}_{k:k+N}, \bm{d}_{k:k+N}, \theta_k) \\
		&x_{k|k} = x_k \\
		&h(\tilde{\bm{y}}_{k:k+N}, \tilde{\bm{u}}_{k:k+N}, \bm{d}_{k:k+N}, \theta_k) \leq 0 \\
		&g(\tilde{\bm{y}}_{k:k+N}, \tilde{\bm{u}}_{k:k+N}, \bm{d}_{k:k+N}, \theta_k) = 0,
	\end{aligned}
\end{equation}
where $x_k$ is the current state of the system and $\bm{d}_{k:k+N}$ is the future disturbance over the prediction window.
Note that this disturbance can be an inaccurate estimate; however, for notational simplicity in the following, we take it to be exact.
The functions $\tilde{f}$ and $\tilde{c}$ constitute a prediction model that approximates the dynamics \eqref{eq:problem_setting-dynamics}, while the parameter $\theta \in \mathbb{R}^{n_\theta}$ contains MPC parameters that define the behavior of the MPC controller, e.g., model weights in a neural network (NN) prediction model approximation for \eqref{eq:problem_setting-dynamics}, or coefficients in the cost function.
The applied control input is the first element of the input optimizer $u(x_k, \bm{d}_{k:k+N}, \theta_k) = \tilde{u}_{k|k}^\ast$.

The performance of the MPC controller \eqref{eq:problem_setting-mpc} depends on the MPC parameters $\theta$, the system parameters $\phi$, and the operating regions of $x$ and $d$ for the closed-loop system.
As disturbances for which the MPC prediction model is inaccurate are encountered, or as $\phi$ changes due to, e.g., slow changes in the underlying physical system, an MPC controller with parameter $\theta$ may experience performance degradation, eventually failing to acceptably control the system.
We consider the problem of monitoring this `acceptability' online, and to adapt $\theta$ such that the controller performance remains acceptable.

\section{Online Monitoring and Adaptation}
This section provides the solution to the above problem.
% In this section we present the online monitoring and adaptation approach for maintaining the performance of the MPC controller \eqref{eq:problem_setting-mpc}.
First, the notion of `acceptability' is formalized, and it is shown how this acceptability can be monitored online.
Second, a performance-based approach for adapting $\theta$ is presented as an initial response to loss of acceptability.
Finally, this is combined with sysID as a backup method for adapting $\theta$, triggered by performance monitoring, in a full scheme for online monitoring and adaptation.

\subsection{Performance Monitoring}\label{sec:monitoring}
As discussed in Section \ref{sec:introduction}, performance monitoring of MPC controllers typically considers either the prediction model \cite{zhao_multi-step_2010, de2024lifelong}, system variable trajectories \cite{alghazzawi_model_2009}, or the incurred cost of the controller \cite{schafer2004multivariable}.
Considering only prediction accuracy or system variables may be overly conservative, as predictions can degrade, or variable trajectories can change, without signifying worsening control performance \cite{piga2019performance}.
% Furthermore, monitoring prediction accuracy is not compatible with the paradigm of performance-based adaptation, in which the adaptation may even worsen prediction quality.
Alternatively, the cost incurred by the controller cannot be monitored as an absolute measure, as it depends on the system parameter $\phi$ and the operating regions for $x$ and $d$.
Furthermore, monitoring only the cost does not consider other modes of degradation, e.g., loss of efficiency in energy systems, or excessive violations of constraints.
In this work, to effectively monitor the performance of the MPC controller, we define acceptability based on a statistical measure of a suite of \emph{features} that can capture general notions of desired performance. 

\textcolor{black}{Let us define a set of $M$ \emph{features} stacked in the vector
\begin{equation}\label{eq:monitor-z_def}
	z_{k, k^\prime} = \bigg[\sigma_{k, k^\prime}^{[1]}, \dots, \sigma_{k, k^\prime}^{[M]}\bigg]^\top,
\end{equation}}
where each feature $\sigma_{k, k^\prime}^{[m]} \in \mathbb{R}$ is defined by
\begin{equation}
	\label{eq:monitoring-features}
	\sigma_{k, k^\prime}^{[m]} = F_m\big(x_k, \bm{d}_{k: k^\prime+N}, \theta_k, \phi \big),
\end{equation} 
i.e., the feature is derived from the closed-loop system \eqref{eq:problem_setting-dynamics}, with system parameter $\phi$, between time steps $k$ and $k^\prime$ under the controller $u = u(x_k, \bm{d}_{k:k+N}, \theta_k)$, such that
\begin{equation}
	x_{k+1} = f\big(x_k, u(x_k, \bm{d}_{k:k+N}, \theta_k), d_k, \phi \big).
\end{equation}
\textcolor{black}{The subscript ${k,k^\prime}$ differs from ${k:k^\prime}$ in that it denotes a scalar value computed between time steps $k$ and $k^\prime$, rather than a stacked vector.} 
It is assumed that the functions $F_m$ are unknown, but that the features $\sigma_{k, k^\prime}^{[m]}$ are observable; examples include average economic costs, settling times, disturbance magnitudes, constraint margins, or variances of controlled variables at steady-state.
% Note that the dependence on the additional $N$ disturbances $\bm{d}_{k: k^\prime+N}$ is due to the control input at time step $k^\prime$ depending on $\bm{d}_{k^\prime: k^\prime+N}$ via \eqref{eq:problem_setting-mpc}.

% In order to perform multi-variate analysis, we introduce the stacked feature vector %$z_{k, k^\prime} = \big[\sigma_{k, k^\prime}^{[1]}, \dots, \sigma_{k, k^\prime}^{[M]}\big]^\top \in \mathbb{R}^L$.
%\iffalse
% \begin{equation}\label{eq:monitor-z_def}
% 	z_{k, k^\prime} = \bigg[\sigma_{k, k^\prime}^{[1]}, \dots, \sigma_{k, k^\prime}^{[M]}\bigg]^\top.
% \end{equation}
%\fi
The following assumption gives the key prerequisite of the proposed method: a baseline data set.
\begin{assumption}
	A baseline data set of feature vectors is available:
	\begin{equation}
		\mathcal{D} = \big\{z_{k, k^\prime}\big\}_{(k, k^\prime) \in \mathcal{K}},
	\end{equation} 
	with $\mathcal{K}$ a set of equidistant pairs of time steps.
	The data set is generated by combinations of the system \eqref{eq:problem_setting-dynamics}, with parameter $\phi$, and an MPC controller \eqref{eq:problem_setting-mpc}, with parameter $\theta$, for which the control performance is considered acceptable.
\end{assumption}
This baseline data set could be generated during a data-collection phase in which the controller is known to perform well, or drawn from historical data.
\textcolor{black}{Importantly, the data is made available \emph{a priori}, and need not be generated online as is required for re-identification with sysID.}
\textcolor{black}{For monitoring, considering each feature individually does not capture how correlations between them describe acceptable performance, 
%e.g., for energy systems, the features of economic cost and power demand are highly correlated, 
and monitoring each individually can lead to false positives for acceptability.}
We then define a statistical distance which is used to measure proximity to this baseline data set.
\begin{definition}
	The statistical Mahalanobis distance $T^2(z, \mathcal{D})$, that computes a distance between the feature vector $z$ and the data set $\mathcal{D}$, is defined as \cite{montgomery2012statistical}
	\begin{equation}
		T^2(z, \mathcal{D}) = (z - \mu_{\mathcal{D}})^\top (\Sigma_{\mathcal{D}})^{-1} (z - \mu_{\mathcal{D}}),
	\end{equation}
	with $\mu_{\mathcal{D}} = \frac{1}{|\mathcal{K}|} \sum_{(k, k^\prime) \in \mathcal{K}} z_{k, k^\prime}$ and
	\begin{equation}
		\begin{aligned}
			\Sigma_{\mathcal{D}} &= \frac{1}{|\mathcal{K}| - 1} \sum_{(k, k^\prime) \in \mathcal{K}} (z_{k, k^\prime} - \mu_{\mathcal{D}}) (z_{k, k^\prime} - \mu_{\mathcal{D}})^\top.
		\end{aligned}
	\end{equation}
\end{definition}
We then define acceptability of the MPC controller \eqref{eq:problem_setting-mpc} as follows.
\begin{definition}
	For a threshold $\alpha > 0$, the set of acceptable MPC parameters $\theta$ is defined as
	\begin{equation}\label{eq:monitoring-acceptability}
		\Theta\big(x_k, \bm{d}_{k: k^\prime+N}, \phi\big) = \big\{\theta \: | \: T^2(z_{k, k^\prime}, \mathcal{D}) \leq \alpha \big\},
	\end{equation}
	where the dependence of $z_{k, k^\prime}$ on $x_k, \bm{d}_{k: k^\prime+N}, \theta$, and $\phi$ is via \eqref{eq:monitor-z_def} and \eqref{eq:monitoring-features}.
\end{definition}
%The selection of the threshold $\alpha$ encodes a certain confidence interval that the the feature vector $z$ is in distribution for the data set $\mathcal{D}$ \cite{montgomery2012statistical}.
The acceptability definition \eqref{eq:monitoring-acceptability} defines the set of MPC parameters that keep the behavior of the controller statistically proximal to that defined by $\mathcal{D}$.
To monitor the acceptability of the MPC controller online, we observe the features $z_{k, k^\prime}$ and monitor the condition $T^2(z_{k, k^\prime}, \mathcal{D}) \leq \alpha$ to check whether the current MPC parameter is acceptable, i.e., $\theta \in \Theta$.

\begin{remark}
	The form of the features $\sigma_{k,k^\prime}^{[m]}$, as well as the time step range over which they are computed, via the choice of $k^\prime$, are design choices on the part of the control designer, and will vary based on the control task, e.g., settling time and overshoot for fast set-point stabilizing controllers, or economic factors for slower economic controllers.
	% Monitoring via \eqref{eq:monitoring-acceptability} allows the designer to define the notion of `acceptability' via careful design of the features.
	We highlight that features related to the state $x$ and disturbance $d$ should be included in the set of features, as the acceptable values of other features will, in general, depend on the current operating region of the system.
    %, e.g., economic cost and load demand disturbances in energy systems.
	Exceptions to this are systems or features that are insensitive to either $x$ or $d$, e.g., the variance of a variable around a set point at steady state is independent of the initial state of the system.
	% Naturally, when features relating to $x$ and $d$ are included, monitoring can only be performed for $x$ and $d$ covered by the features in $\mathcal{D}$.
\end{remark}

\subsection{Performance-Based Adaptation}\label{sec:online_adaptation}
We now explore how acceptability can be restored online using performance-based learning.
The MPC parameter $\theta$ is modified online towards the set $\Theta$, thus aiming to reduce the statistical measure $T^2(z_{k, k^\prime}, \mathcal{D})$ below the threshold $\alpha$.
%Note that, in theory, any learning algorithm that uses $T^2(z_{k, k^\prime}, \mathcal{D})$ as a cost function to be minimized can be applied to adapt $\theta$.
In this work, we propose the use of MPC-based Q-learning: a powerful, online learning method for MPC with relatively few tuning hyper-parameters \cite{gros_data-driven_2020}.

In MPC-based Q-learning, for a control design task
\begin{equation}\label{eq:performance_cost}
	\min_{\pi} \sum_{k=0}^K C_k,
\end{equation} 
with $K \in \mathbb{Z}^+ \cup \{\infty\}, \pi$ the controller, and $C_k$ a stage cost, the MPC optimization problem \eqref{eq:problem_setting-mpc} serves as both controller, with $\pi_k = u(x_k, \bm{d}_{k:k+N}, \theta_k)$ the control action at time step $k$, and value function $V_k = J(x_k, \bm{d}_{k:k+N}, \theta_k)$, an estimate of the future cost given the current state of the system.
The theoretical foundations of the method lie in a key result, first presented in \cite{gros_data-driven_2020}, that states that, by tuning a parameterized MPC cost function, an optimal controller can be found even with an incorrect prediction model.
In practice, the prediction model and constraints are usually parameterized as well, increasing the number of degrees of freedom for discovering an effective policy.

Inspired by this, we split the cost and constraint terms of \eqref{eq:problem_setting-mpc} into parameterized and non-parameterized components.
Furthermore, we specify the MPC parameter to be composed of a prediction model parameter $\tilde{\theta}$ and a controller-tuning parameter $\hat{\theta}$, such that $\theta = \big[\tilde{\theta}^\top, \hat{\theta}^\top\big]^\top$, and the MPC controller becomes:
% introduce additional terms $\hat{L}$, $\hat{h}$, and $\hat{g}$ to the cost and constraints of \eqref{eq:problem_setting-mpc}, parameterized by $\hat{\theta} \in \mathbb{R}^{n_{\hat{\theta}}}$, such that the MPC controller becomes
\begin{equation}\label{eq:adaptation-mpc}
	\begin{aligned}
		J\big(x_k, &\bm{d}_{k:k+N}, \theta_k\big) \\
		&= \min_{\substack{\tilde{\bm{x}}_{k:k+N+1},\\ \tilde{\bm{y}}_{k:k+N}, \\ \tilde{\bm{u}}_{k:k+N}}} \: L(\tilde{\bm{x}}_{k:k+N+1}, \tilde{\bm{y}}_{k:k+N}, \tilde{\bm{u}}_{k:k+N}, \bm{d}_{k:k+N}) \\
		&\quad \quad \quad \quad + \hat{L}(\tilde{\bm{x}}_{k:k+N+1}, \tilde{\bm{y}}_{k:k+N}, \tilde{\bm{u}}_{k:k+N}, \bm{d}_{k:k+N}, \hat{\theta}_k) \\
		\text{s.t.} \quad &\tilde{\bm{x}}_{k+1:k+N+1} = \tilde{f}(x_k, \tilde{\bm{u}}_{k:k+N}, \bm{d}_{k:k+N}, \tilde{\theta}_k)  \\
		&\tilde{\bm{y}}_{k:k+N} = \tilde{c}(\tilde{\bm{x}}_{k:k+N}, \tilde{\bm{u}}_{k:k+N}, \bm{d}_{k:k+N}, \tilde{\theta}_k) \\
		&x_{k|k} = x_k \\
		&h(\tilde{\bm{y}}_{k:k+N}, \tilde{\bm{u}}_{k:k+N}, \bm{d}_{k:k+N}) \\
		 &+ \hat{h}(\tilde{\bm{y}}_{k:k+N}, \tilde{\bm{u}}_{k:k+N}, \bm{d}_{k:k+N}, \hat{\theta}_k) \leq 0 \\
		&g(\tilde{\bm{y}}_{k:k+N}, \tilde{\bm{u}}_{k:k+N}, \bm{d}_{k:k+N}) \\
		 &+ \hat{g}(\tilde{\bm{y}}_{k:k+N}, \tilde{\bm{u}}_{k:k+N}, \bm{d}_{k:k+N}, \hat{\theta}_k) = 0,
	\end{aligned}
\end{equation}
where the tuning terms are equal to zero for $\hat{\theta}=0$, e.g., $\hat{L}(\cdot, \cdot, \cdot, \bm{0}) = 0$.
The non-parameterized cost and constraints represent nominal MPC controller components, e.g., an economical cost with known prices, or constraints as known physical limits, which function well when the prediction model parameters $\tilde{\theta}$ result in accurate predictions.
The additional terms, the structures of which are to be designed by the control designer, are additional degrees of freedom that can aid in restoring performance when the prediction model parameters $\tilde{\theta}$ result in inaccurate predictions.
% These additional terms, provide additional degrees of freedom to improve the performance of the MPC controller.
The parameters are then updated online\footnote{For notational simplicity, we have provided the update in terms of value functions with recursive updates, which holds when no exploration or experience buffers are included in the RL algorithm. See \cite{gros_data-driven_2020} for a more general version.}, via the reinforcement learning-based approach from \cite{gros_data-driven_2020}, as
\begin{equation}\label{eq:adaptation-update}
	\theta_{k+1} = \theta_k + \beta \delta_k \nabla_{\theta} J\big(x_k, \bm{d}_{k:k+N}, \theta_k\big)
\end{equation}
with $\beta > 0$ a learning rate, and where
\begin{equation*}
	\begin{aligned}
		\delta_k &= C_k + \gamma J(x_{k+1}, \bm{d}_{k+1:k+N+1}, \theta_k)  - J(x_{k}, \bm{d}_{k:k+N}, \theta_k),
	\end{aligned}
\end{equation*}
with the sensitivity $\nabla_{\theta} J$ available automatically upon solving \eqref{eq:adaptation-mpc}; see \cite{gros_data-driven_2020} for more details.

%\textcolor{black}{Note that, the performance-based adaptation \eqref{eq:adaptation-update} adapts the controller to minimize the sum of $L_k$, and is not therefore explicitly pushing $\theta$ to $\Theta$.
%The cost $L_k$ can be considered a proxy cost.
%Future work will look at leveraging learning algorithms that minimize explicitly $T^2$,}

\subsection{High-Level Flow}
%\begin{figure}
%	\centering
%	\includegraphics[scale=0.65]{media/ink/flow_chart.pdf}
%	\caption{Online monitoring and adaptation scheme.}
%	\label{fig:flow}
%\end{figure}
\begin{algorithm}
\small
    \caption{Online monitoring and adaptation scheme.}\label{alg:scheme}
    \begin{algorithmic}[1]
        \While{true}
            \State Observe $z_{k,k^\prime}$
            \If{$T^2(z_{k,k^\prime}, \mathcal{D}) > \alpha$}
                \State Update $\theta$ as \eqref{eq:adaptation-update}
                \State Observe $z_{k,k^\prime}$
                \If {$T^2(z_{k,k^\prime}, \mathcal{D}) > \alpha$}
                    \State Update $\tilde{\theta}$ via sysID
                    \State Reset $\hat{\theta}$ to zero
                \EndIf
            \EndIf
        \EndWhile
    \end{algorithmic}
\end{algorithm}
We now propose a scheme, presented in Algorithm \ref{alg:scheme}, that uses the performance-based controller adaptation \eqref{eq:adaptation-update} as a primary response to loss of acceptability, and a traditional sysID approach as a fallback mechanism.
The feature vector $z_{k,k^\prime}$ is observed continuously and, if \eqref{eq:monitoring-acceptability} indicates $\theta \notin \Theta$, the performance-based learning phase \eqref{eq:adaptation-update} is triggered; if this fails to shift $\theta$ to $\Theta$, i.e., $T^2$ remains larger than $\alpha$, a sysID update of $\tilde{\theta}$ is performed, with $\hat{\theta}$ reset to zero.

This scheme is motivated by the attitude that sysID is, in general, the most robust approach to restoring performance, but may require significant disturbance, or even temporary disconnection, of the system to generate highly exciting data, and is thus undesirable as a response to slight, and potentially frequent, degradation.
In contrast, performance-based learning can restore acceptability quickly online, leveraging the additional degrees of freedom in \eqref{eq:adaptation-mpc}.
Note that this triggered use of performance-based learning contrasts with its typical continuous application \cite{gros_data-driven_2020, sorourifar2021data}, here being switched on and off to maintain $\theta \in \Theta$.

\begin{remark}
    While, for simplicity, Algorithm \ref{alg:scheme} transitions to adaptation immediately upon the acceptability threshold being breached, it may be desirable, in practice, to monitor for persistent breaching of this threshold before triggering adaptation.
    \textcolor{black}{Likewise, the same holds for transitioning out of adaptation or deploying sysID.}
\end{remark}

\section{Case Study}
\begin{figure}[t!]
	\centering
	\includegraphics[width=0.3 \textwidth]{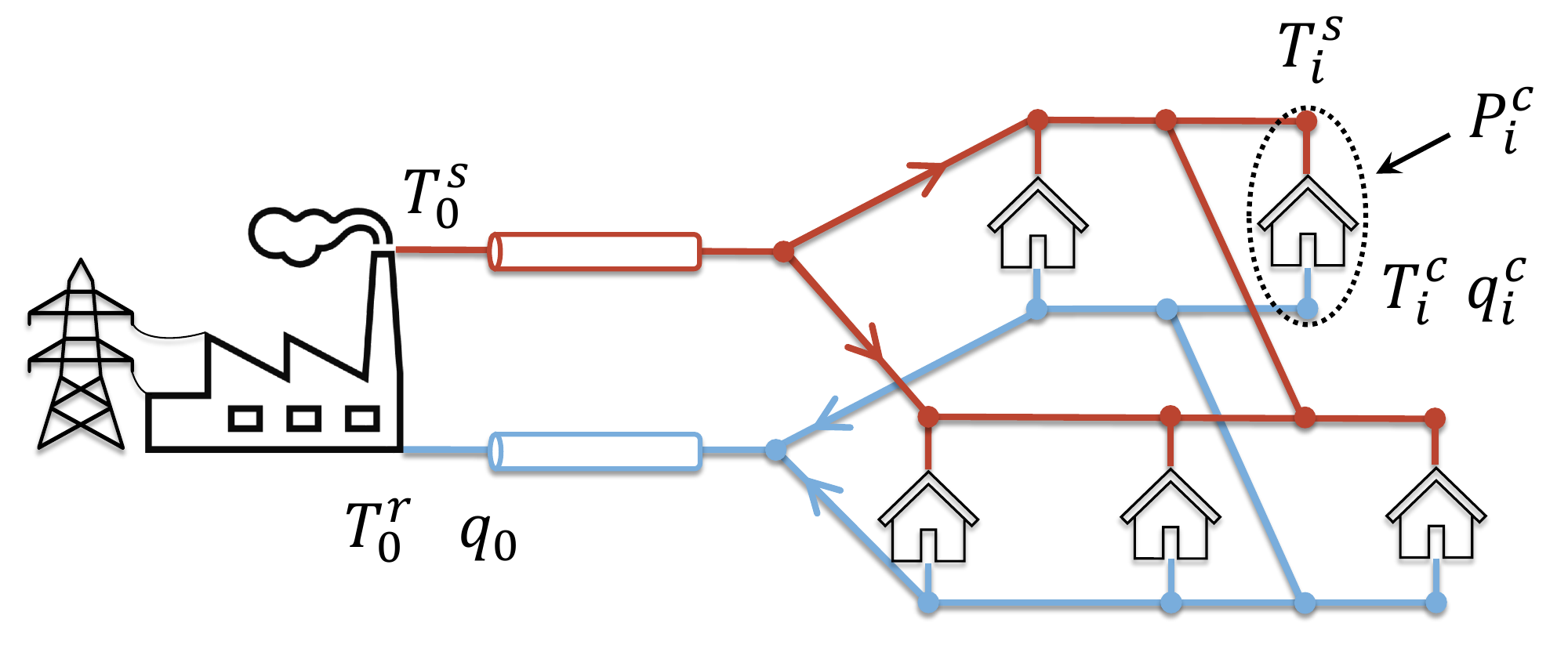}
	\caption{Schematic of the AROMA DHS and its main variables.}
	\label{fig:aroma}
\end{figure}

The proposed approach is demonstrated on a district heating system (DHS), a large-scale energy system that distributes heat in a closed network between heat sources and consumers. 
% A DHS typically consists of a heating station with multiple thermal generators and an insulated water pipeline network that transfers heat to thermal loads.
% These loads use local heat exchangers to absorb the delivered heat for indoor heating and domestic hot water  \cite{la2023optimal}.
The specific case study analyzed in this work considers the AROMA DHS, presented in \cite{krug2021nonlinear}, depicted in Figure \ref{fig:aroma}, and simulated with a high-fidelity dynamic model based on a dedicated Modelica library \cite{nigro2024control}. Optimization problems are constructed using Casadi and solved using Ipopt.
Python source code for the following is available at \url{https://github.com/SamuelMallick/mpcrl-process}.

The following inputs, outputs, and disturbances are considered, as shown in Figure \ref{fig:aroma} and described in \cite{de2024physics}. 
The manipulated (control) variable is the supply temperature at the heating station, denoted by \mbox{$u_k=T_{0,k}^\text{s}$}.
The output variable is \mbox{$y_k \!=\! [ T_{0,k}^\text{r}, q_{0,k},  T_{1,k}^\text{s},  \hdots,  T_{5, k}^\text{s}, T_{1,k}^\text{c},  \hdots,  T_{5,k}^\text{c}, q_{1,k}^\text{c},  \hdots,  q_{5,k}^\text{c}]^{\top}$}, with $T_0^\text{r}$ and $q_0$ the temperature and water flow, respectively, at the heating station, and with $T_i^\text{s}$ the supply temperature, $T_i^\text{c}$ the output temperature, and $q_i^\text{c}$ the water flow, for each $i$-th thermal load.
The disturbance $d_k = [P_{1, k}^\text{c}, \dots, P_{5, k}^\text{c}, c_{\text{elec}, k}, \underline{T}_{0, k}^\text{r}, \underline{T}_k^\text{s}]^\top$ includes the five thermal load demands $P_i^\text{c}$ for $i=1,\dots,5$, the electricity price $c_\text{elec}$, and the lower bounds on return $\underline{T}_0^\text{r}$ and supply $\underline{T}^\text{s}$ temperatures. 
Note that all temperatures are expressed in [$^\circ$C], all water flow rates in [kg/s], and all powers in [W]. 
% Overall, the system has $n_u = 1$ inputs,  $n_y = 17$ outputs, $n_d = 8$ disturbances, and 
%The system is governed by highly non-linear thermal and transportation  dynamics across many components, e.g., pipes and heat exchangers, making it a challenging system to accurately model with physical laws.

The system is controlled by an economic MPC controller that minimizes the cost of generating power while satisfying load power consumption and system constraints on variables.
The MPC controller \eqref{eq:adaptation-mpc} is implemented with horizon $N = 72$, a sampling time of 300s (5 minutes), and economic cost
\begin{equation}
	\begin{aligned}
		L(&{\bm{y}}_{k:k+N}, \tilde{\bm{u}}_{k:k+N}, \bm{d}_{k:k+N}) = \\
		 &\sum_{\tau=0}^{N} \ell\big(\tilde{q}_{0, k+\tau|k}, \overline{T}_{0, k+\tau|k}^\text{s}, \overline{T}_{0, k+\tau|k}^\text{r}, c_{\text{elec}, k+\tau}\big),
	\end{aligned}
\end{equation}
where $\ell(q_{0}, T_{0}^\text{s}, T_{0}^\text{r}, c_{\text{elec}}) = \tau \cdot c_{\text{elec}} \cdot P_\text{b}(q_0, T_0^\text{s}, T_0^\text{r})$,
\iffalse
\begin{equation}
	\ell(q_{0}, T_{0}^\text{s}, T_{0}^\text{r}, c_{\text{elec}}) = \tau \cdot c_{\text{elec}} \cdot P_\text{b}(q_0, T_0^\text{s}, T_0^\text{r})
\end{equation}
\fi
and $P_\text{b}(q_0, T_0^\text{s}, T_0^\text{r}) = c_\text{p} \cdot q_0 \cdot \big(T_0^\text{s} - T_0^\text{r}\big)$ is the generated power with $c_\text{p}$ the specific water heat coefficient.
Furthermore, the constraint $h$ stacks constraints over the prediction window, i.e., $h = [h_{k|k}(\cdot)^\top, \dots, h_{k+N|k}(\cdot)^\top]^\top \leq 0$ with each element containing bounds on inputs and outputs\footnote{Note that the output constraints are `softened' with slack variables that are penalized in the cost to ensure feasibility.}
   \begin{equation*}
\small
	\begin{aligned}
		&h_{k+i|k}\big(\tilde{y}_{k|k+i}, \tilde{u}_{k|k+i}, \underline{T}_{0,k+i}^\text{r}, \underline{T}^\text{s}_{k+i} \big)=\\ & [
		10 - \tilde{q}_0, 
		\tilde{q}_0 - 25, 
		\overline{T}_0^\text{r} - 75,
		\underline{T}_{0}^\text{r} - \overline{T}_0^\text{r},\overline{T}_1^\text{s} - 85, \\
		&\quad \;\; 
		\underline{T}^\text{s} -  \overline{T}_1^\text{s}, 
		\dots, 
		\overline{T}_5^\text{s} - 85, \underline{T}^\text{s} -  \overline{T}_5^\text{s}, 
		\overline{T}_0^\text{s} - 85, 
		65 - \overline{T}_0^\text{s}
		]^\top\\&\quad \;\; \forall i \in \{1,\hdots,N\}
	\end{aligned}
\end{equation*}% $h = [h_\text{k}(\cdot), \dots, h_\text{k}(\cdot)]^\top \leq 0$,
\iffalse
\begin{equation}
	\begin{aligned}
		h(&\tilde{\bm{y}}_{k:k+N}, \tilde{\bm{u}}_{k:k+N}, \bm{d}_{k:k+N}) = \\
		 &\begin{bmatrix}
			h_\text{k}\big(\tilde{y}_{k|k}, \tilde{u}_{k|k}, \underline{T}_{0, k}^\text{r}, \underline{T}_k^\text{s} \big) \\
			\vdots \\
			h_\text{k}\big(\tilde{y}_{k+N|k}, \tilde{u}_{k+N|k}, \underline{T}_{0, k+N}^\text{r}, \underline{T}_{k+N}^\text{s} \big)
		\end{bmatrix} \leq 0,
	\end{aligned}
\end{equation}
\begin{equation*}
\small
	\begin{aligned}
		&h_\text{k}\big(\tilde{y}, \tilde{u}, \underline{T}_{0}^\text{r}, \underline{T}^\text{s} \big) = [
		10 - \tilde{q}_0, 
		\tilde{q}_0 - 25, 
		\overline{T}_0^\text{r} - 75,
		\underline{T}_{0}^\text{r} - \overline{T}_0^\text{r}, \\
		&\overline{T}_1^\text{s} - 85,
		\underline{T}^\text{s} -  \overline{T}_1^\text{s}, 
		\dots, 
		\overline{T}_5^\text{s} - 85, \underline{T}^\text{s} -  \overline{T}_5^\text{s}, 
		\overline{T}_0^\text{s} - 85, 
		65 - \overline{T}_0^\text{s}
		]^\top.
	\end{aligned}
\end{equation*}
\fi
% There are no equality constraints.
Furthermore, for the prediction model, the data-based model proposed in \cite{de2024physics} is used, where \eqref{eq:problem_setting-dynamics} is represented by a recurrent NN using gated recurrent units, and where $\tilde{\theta}$ contains the network weights.
The model is identified in a sysID fashion offline; see \cite{de2024physics} for details.
Finally, the additional terms in \eqref{eq:adaptation-mpc} are 
\begin{equation*}
	\begin{aligned}
		\hat{L}(&\tilde{\bm{y}}_{k:k+N}, \tilde{\bm{u}}_{k:k+N}, \bm{d}_{k:k+N}, \hat{\theta}) = \\
		 &\sum_{\tau = 0}^N \Big( f_{\hat{\theta}}^\top v_{k+\tau} + v_{k+\tau}^\top Q_{\hat{\theta}} v_{k+\tau} \Big) + \omega_{\hat{\theta}} \sum_{i=1}^5 (\overline{T}_{i, k+N}^\text{s} - T_{\hat{\theta}})^2,
	\end{aligned}
\end{equation*}
with $v_{k} = \Big[\tilde{u}_{k|k}^\top, \tilde{y}_{k|k}^\top, P_\text{b}\big(\tilde{q}_{0, k|k}, \overline{T}_{0, k}^\text{s}, \overline{T}_{0, k}^\text{r}\big)\Big]^\top$, and
\begin{equation*}
	\begin{aligned}
		\hat{h}(&\tilde{\bm{y}}_{k:k+N}, \tilde{\bm{u}}_{k:k+N}, \bm{d}_{k:k+N}, \hat{\theta}) = \\
		&[\Delta q_{\hat{\theta}},\Delta q_{\hat{\theta}}, 0, 0, \Delta T_{\hat{\theta}}, 
		\dots, \Delta T_{\hat{\theta}}, 0, 0]^\top.
	\end{aligned}
\end{equation*}
These terms include linear and quadratic penalties on input and output variables, such that the controller can learn to avoid extreme values, a terminal temperature penalty, such that the controller can learn to maintain a certain energy in the system, and constraint back-off parameters, such that the controller can learn to be conservative with respect to constraints.
Thus, the full parameter is $\hat{\theta} = [f_{\hat{\theta}}^\top, \tilde{Q}_{\hat{\theta}}^\top, \omega_{\hat{\theta}}, T_{\hat{\theta}}, \Delta q_{\hat{\theta}}, \Delta T_{\hat{\theta}}]^\top$,
\iffalse
\begin{equation}
	\hat{\theta} = [f_{\hat{\theta}}^\top, \tilde{Q}_{\hat{\theta}}^\top, \omega_{\hat{\theta}}, T_{\hat{\theta}}, \Delta q_{\hat{\theta}}, \Delta T_{\hat{\theta}}]^\top,
\end{equation}
\fi
where $\tilde{Q}_{\hat{\theta}}$ contains the entries of the matrix $Q_{\hat{\theta}}$, flattened into a vector.

For performance monitoring we consider 8 features computed over windows of 12 hours (144 time steps) $z_{k,k+144} = [\sigma_{k,k+144}^{[1]}, \dots, \sigma_{k,k+144}^{[8]}]^\top$ with
\begin{subequations}
    \begin{align}
        \sigma_{k,k+144}^{[i]} &= \frac{1}{144} \sum_{\tau = k}^{k+144} \phi_{i, k}, \: i=1,\dots,4, \label{eq:feature_mean} \\
        \sigma_{k,k+144}^{[i]} &= \frac{1}{143} \sum_{\tau = k}^{k+144} (\phi_{i-4, k} - \sigma_{k,k+144}^{[i-4]})^2, \:i=5,\dots,8, \label{eq:feature_var}
    \end{align}
\end{subequations}
and where 
\begin{subequations}
	\begin{align}
		\phi_{1, k} &= P_\text{b}(q_{0, k}, T_{0, k}^\text{s}, T_{0, k}^\text{r}) / \sum_{i=1}^5 |P_{i, k}^\text{c}| \label{eq:feature_eff} \\
		\phi_{2, k} &= \tau \cdot c_{\text{elec}, k } \cdot P_\text{b}(q_{0, k}, T_{0, k}^\text{s}, T_{0, k}^\text{r}) \label{eq:feature_cost} \\
		\phi_{3, k} &= \zeta\big(q_{0}, \underline{T}^\text{s}, T_{1, k}^\text{s}, \dots,T_{5, k}^\text{s} \big) \label{eq:feature_viol}\\
		\phi_{4, k} &= \sum_{i=1}^5 |P_{i, k}^\text{c}|, \label{eq:feature_load}
	\end{align}
\end{subequations}
with $\zeta\big(q_{0}, \underline{T}^\text{s}, T_{1, k}^\text{s}, \dots,T_{5, k}^\text{s}\big) = \max(0, 10 - q_{0}) \textcolor{black}{+ \max(0, q_{0} - 25)} + \sum_{i=1}^5 \max(0, \underline{T}^\text{s} - T_{i}^\text{s})$.
\iffalse
\begin{equation}
	\begin{aligned}
		\zeta\big(q_{0}, \underline{T}^\text{s}, T_{1, k}^\text{s}, \dots,T_{5, k}^\text{s}\big) &= \max(0, 10 - q_{0}) \\
		 &\quad+ \sum_{i=1}^5 \max(0, \underline{T}^\text{s} - T_{i}^\text{s}).
	\end{aligned}
\end{equation}
\fi
These features represent the mean \eqref{eq:feature_mean} and variance \eqref{eq:feature_var} of the system efficiency \eqref{eq:feature_eff}, the economic cost of power generation \eqref{eq:feature_cost}, the violation of output constraints \eqref{eq:feature_viol}, and the total load demand \eqref{eq:feature_load}.

The data set $\mathcal{D}$ is generated from 35 days of operation where the MPC controller (with all additional parameters $\hat{\theta}$ zero) operates acceptably.
%Representative figures are provided in the longer version of this article online, and exact values can be found in the provided code base.
%\textcolor{black}{This is achieved through the use of an extremely able model, which is unrealistic.}
% Three representative days of disturbances during the data generation are shown in Figure \ref{fig:disturbs}.
The load powers are generated randomly for each day as $P_i^\text{c} = a \cdot \hat{P}_i$, with $a$ uniformly sampled from $[0.6, 1.4]$, where $\hat{P}_i$ is a nominal load profile representing a typical profile observed in DHSs.
%(first day in Figure \ref{fig:disturbs}) \cite{la2023optimal}.
Similarly, the electricity price and temperature bounds represent typical behavior, with electricity prices peaking in the morning and the evening, and with lower temperatures allowed at night.
Exact values for all disturbances can be found in the source code.

\iffalse
\begin{figure}
	\centering
	\input{media/tikz/disturb}
	\caption{Representative load disturbances (top), electricity price (middle), and lower bound for return (dashed) and supply (solid) temperatures (bottom) during data generation.}
	\label{fig:disturbs}
\end{figure}
\fi
The features $\sigma_{k,k+144}^{[m]}$ are then sampled from this operational data, with $k$ sampled randomly with uniform probability over the 35 days. 
As threshold $\alpha$ we take $\alpha=15.51$, which corresponds to a $95\%$ confidence interval for the observed $z_{k,k^\prime}$ to be within the distribution defined by $\mathcal{D}$ \cite{montgomery2012statistical}\footnote{For this confidence interval to hold, the data set $\mathcal{D}$ must be multivariate normal. As this is not strictly satisfied, this choice can be viewed as an informed heuristic.}.
%\begin{remark}
	%A common statistical approach to selecting $\alpha$ assumes that $T^2$ is characterized by a chi square distribution, allowing $\alpha$ to define an ellipsoidal region, such that all points within the region are statistically likely, given $\mathcal{D}$, with a certain confidence  \cite{montgomery2012statistical}, e.g., for $L=8$, $T^2 < 15.51$ defines a region of $95\%$ confidence.
%\end{remark}
For performance-based adaptation in \eqref{eq:performance_cost} we use
\begin{equation*}
	C_k = \ell(q_{0, k}, T_{0, k}^\text{s}, T_{0, k}^\text{r}, c_{\text{elec}, k}) + \zeta\big(q_{0}, \underline{T}^\text{s}, T_{1, k}^\text{s}, \dots,T_{5, k}^\text{s}\big),
\end{equation*}
and learning rate $\beta = 0.1$.
\textcolor{black}{We update only the parameters $\hat{\theta}$ with the Q-learning update \eqref{eq:adaptation-update}}, demonstrating how the extra degrees of freedom in \eqref{eq:adaptation-mpc} can be sufficient for maintaining acceptability.

We demonstrate three cases in the following experiments:

\textbf{Case 1}: A change in $\phi$.
	In particular, the control input $u = T_0^\text{s}$ applied to the system has a negative offset of one degree $^\circ$C with respect to the setpoint requested by the controller. This can happen in case of unmodeled losses in the system.
	
\textbf{Case 2}: The load demand transitions to an operating range that was not present in the data used to identify the prediction model, such that the prediction model with parameter $\tilde{\theta}$ is slightly inaccurate.
	See the provided code for exact values.
	
\textbf{Case 3}: The load demand profile transitions as in case 2, however to a larger extent, such that the prediction model with parameter $\tilde{\theta}$ is significantly inaccurate.
\iffalse
\begin{itemize}
	\item \textbf{Case 1}: The underlying system undergoes a structural change, i.e, a change in $\phi$.
	In particular, the control input $u = T_0^\text{s}$ applied to the system has a negative offset of one degree $^\circ$C with respect to the setpoint requested by the controller. This can happen in case of unmodeled losses in the system.
	
	\item \textbf{Case 2}: The load demand profile transitions to an operating range that was not present in the data used to identify the prediction model, such that the prediction model with parameter $\tilde{\theta}$ is slightly inaccurate.
	See the provided code base for exact values.
	
	\item \textbf{Case 3}: The load demand profile transitions as in case 2, however to a larger extent, such that the prediction model with parameter $\tilde{\theta}$ is significantly inaccurate.
\end{itemize} 
\fi

Note that, for demonstration, the change in each case occurs after three days, and three days of persistent violation of the $T^2$ threshold are required before adaptation is triggered.
\textcolor{black}{Q-learning adaptation \eqref{eq:adaptation-update} is applied until the updates cease to lower the $T^2$ distance, upon which, if the threshold $\alpha$ continues to be violated, sysID is deployed.}

\begin{figure*}
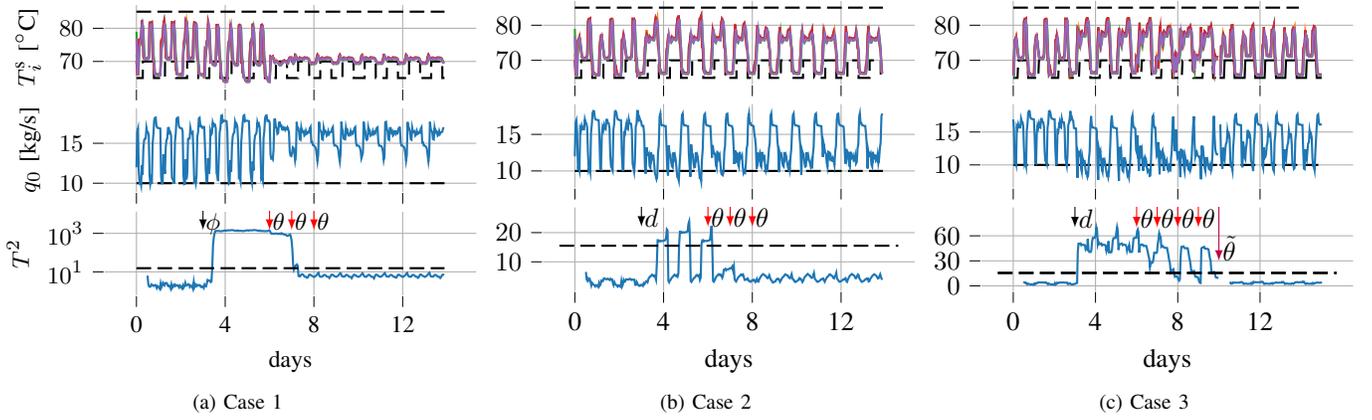

	\centering
    \subfloat[Case 1]{\input{media/tikz/case_1_traj}}
    \subfloat[Case 2]{\input{media/tikz/case_2_traj}}
    \subfloat[Case 3]{\input{media/tikz/case_3_traj}}
	\caption{Time series trajectories of key variables for cases 1, 2, and 3. Time instants where changing conditions and parameter updates occur are indicated with arrows in the bottom plot. The long purple arrow in case 3 indicates when sysID is performed.}
	\label{fig:case_1_traj}
\end{figure*}
\iffalse
\begin{figure}
	\centering
	\input{media/tikz/case_2_traj}
	\caption{Time series trajectories of key variables for case 2. Time instants where changing conditions and parameter updates occur are indicated with arrows in the bottom plot.}
	\label{fig:case_2_traj}
\end{figure}
\begin{figure}
	\centering
	\input{media/tikz/case_3_traj}
	\caption{Time series trajectories of key variables for case 3. Time instants where changing conditions and parameter updates occur are indicated with arrows in the bottom plot. The final time instant (long purple arrow), at the tenth day, indicates when the system is disconnected and sysID is performed.}
	\label{fig:case_3_traj}
\end{figure}
\fi
Figures \ref{fig:case_1_traj} shows the load supply temperatures $T_i^\text{s}$, the return water flow $q_0$, and the statistical distance $T^2$ for cases 1, 2, and 3.
It can be seen that from the third day, when the respective changes come into effect, the controller is unable to produce acceptable behavior, with the statistical distance exceeding the threshold.
As in Algorithm \ref{alg:scheme}, this triggers the performance-based learning response, $\hat{\theta}$ is adjusted, and, for case 1 and 2, acceptable behavior is recovered, with $T^2$, again, consistently below the threshold after eight days.
In particular, the controller introduces relative levels of conservative behavior, restricting temperature oscillations, to avoid violating constraints, without overly sacrificing efficiency or economic cost.
Conversely, for case 3, it can be seen that the performance-based learning response is not able to drive $\theta$ to the set $\Theta$ by adjusting $\hat{\theta}$.
As in Algorithm \ref{alg:scheme}, the sysID response is therefore triggered to adjust $\tilde{\theta}$, with the additional parameters $\hat{\theta}$ reset to zero.
The resulting performance, with the updated prediction model, is, again, acceptable.

\begin{figure}\label{fig:features-case2}
	\centering
	% \subfloat[Case 1]{\input{media/tikz/case_1_features}}
	\subfloat{\input{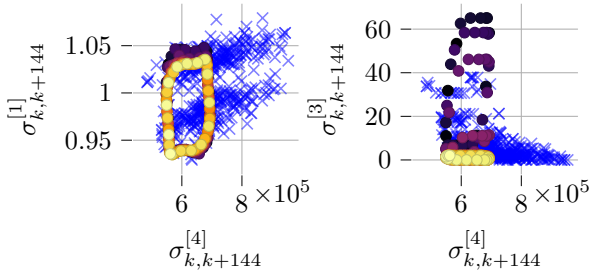}}
	\caption{Slices of feature space for case 2. Blue crosses are data points in $\mathcal{D}$.
	Circles show moving averages of features over 40 time steps after adaptation has been triggered (black for early data, transitioning to yellow for new data).} % monitored online from the third day of operation.
	% In particular, moving averages of features over 40 time steps are shown, with the color transitioning from black, for the earliest data, to light yellow, for the latest data.}
	\label{fig:features}
\end{figure}

While visualizing the statistical distance is not possible in 8 dimensions, Figure \ref{fig:features} shows representative slices for case 2.
It can be seen that the constraint violation feature is pushed to an acceptable range, incurring offsets in other features, e.g., a slight loss of efficiency, however, maintaining the overall statistical distance within the acceptable range.

\section{Conclusions}
This work has presented a novel integrated online monitoring and adaptation strategy for process MPC controllers.
A statistical definition of acceptable performance is introduced as a method of online performance monitoring.
Then, an adaptation scheme that combines performance-based adaptation and system identification is presented, with performance monitoring triggering the different adaptation methods.
The approach was demonstrated in three numerical experiments on a high-fidelity simulator of a district heating network, demonstrating the approach's ability to detect and respond to controller performance degradation.
Future work will look at automatic feature selection, removing the design challenge of feature crafting.

\bibliographystyle{plain}
\bibliography{root.bib}

\end{document}